\documentclass[showpacs,prb,twocolumn,numbers,amsmath]{revtex4}

\usepackage{graphicx}
\usepackage{dcolumn}
\usepackage{bm}
\usepackage{epsfig}
\usepackage{wrapfig}

\begin{document}

\title{Magnetic Compton profiles of Fe and Ni corrected by dynamical
electron correlations}

\author{D. Benea$^{a}$, J. Min\'ar$^{b}$, S. Mankovsky$^{b}$, L. Chioncel$^{c,d}$ and H. Ebert$^{b}$}

\affiliation{$^{a}$Faculty of Physics, Babes-Bolyai University,
Kogalniceanustr 1, Ro-400084 Cluj-Napoca, Romania} 
\affiliation{$^{b}$Chemistry Department, University Munich,
Butenandstr.~5-13, D-81377 M\"unchen, Germany} 
 \affiliation{$^{c}$ Augsburg Center for Innovative Technologies, University of Augsburg, 
D-86135 Augsburg, Germany}
\affiliation{$^{d}$ Theoretical Physics III, Center for Electronic
  Correlations and Magnetism, Institute of Physics, University of
  Augsburg, D-86135 Augsburg, Germany}

\begin{abstract}
Magnetic Compton profiles (MCPs) of Ni and Fe along [111] direction have been
calculated using a combined Density Functional and many-body theory approach. 
At the level of the local spin density approximation 
the theoretical MCPs does not describe correctly the experimental
results around the zero momentum transfer. In this work we demonstrate that
inclusion of electronic correlations as captured by Dynamical Mean Field
Theory (DMFT) improves significantly the agreement between the theoretical
and the experimental MCPs. In particular, an energy decomposition
of Ni MCPs gives indication of spin polarization and intrinsic nature of Ni 6
eV satellite, a genuine many-body feature. 
\end{abstract}


\maketitle

\begin{figure}[b]
\vspace{0.5cm}
\begin{center}
\hspace*{0.2cm}
   \includegraphics[width=0.79\linewidth, clip=true]{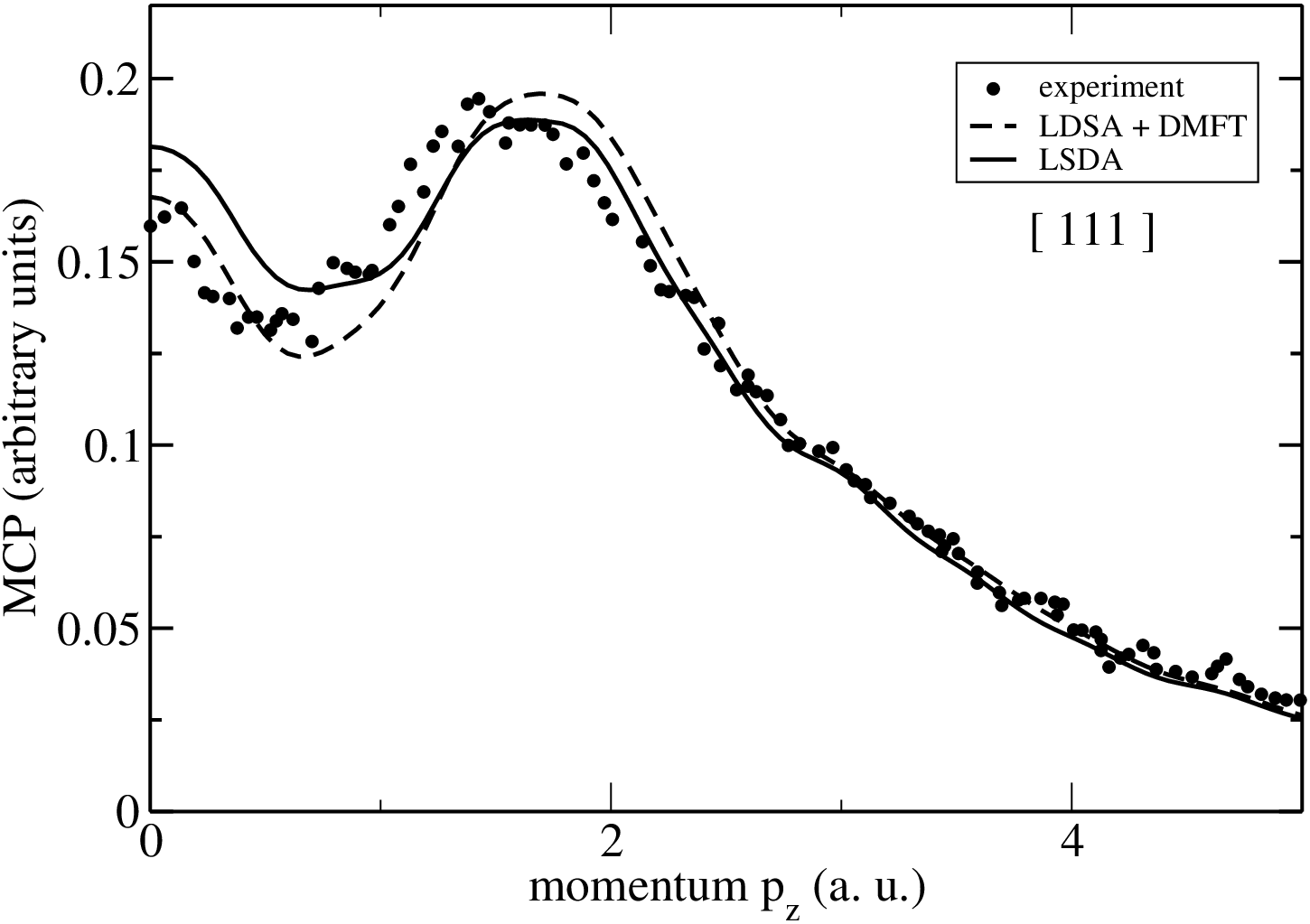}
\end{center}
\caption {\label{Fig:figure1} MCPs of Ni [111] calculated via the KKR method within the LSDA and LSDA+DMFT approach. The theoretical curves have been
convoluted with a Gaussian of 0.43 a.u. FWHM, corresponding to the experimental resolution. The experimental data stem from Dixon 
et al. \cite{DDG+98}.}
\end{figure}

The magnetic Compton scattering is a well-established technique for
probing the spin-dependent momentum densities of magnetic solids
\cite{C85, LC96}. Compared with other experimental techniques, Compton scattering offers
several advantages. Compton scattering is an inelastic process, in
which an energetic photon collides with a single electron and transfers
energy to it. Since the scattering is from a single-electron and (to a
good approximation) occurs at a single point in space, the process
must be incoherent and is supplying
 an average over real space. Therefore, Compton scattering 
is related directly to the electronic
ground state, whereas other spectroscopic methods (e.g. photoemission
spectroscopy) involve excited states. \\
 In addition Compton scattering allows for a rather fundamental test of the theories 
 used to describe the spin dependent momentum density, since these
 theoretical methods are tailored to give predictions for the ground
 state properties.  Several theoretical methods have been used in the past to
 describe the electron momentum density and to ana\-lyze 
the experimental MCPs
 \cite{WK77, KA90, TS93, BZK00, DDG+98, TM06,
 MDW+04}. Most of
the corresponding calculations were done within the local spin density 
approximation (LSDA)
for the exchange-correlation potential. 
In general, the theoretical profiles obtained using LSDA show an
 overall agreement with the experimental measurements, except within
 the region $p_z < 1$ $ a.u.$. It was pointed out that the
 discrepancy between the experimental and theoretical MCP could be
 attributed to missing electron-electron correlations in band
 models \cite{BS85}. Along this line, recently,
 several theoretical methods beyond LSDA have been applied in order 
to describe the features of the MCP which cannot be explained using the LSDA-based approach 
and to improve the agreement with experiment \cite{TM06, K04, SDI+04}. 
The LSDA + U method has been applied by Tokii et al. \cite{TM06} 
to calculate the MCPs of Fe, 
showing improvement for [100] and [110]
directions but still underestimates near the origin the 
MCP for [111] direction. Also, one should mention the calculations 
of the Ni MCP done by Kubo \cite{K04} implemented within 
the GW scheme based on the FLAPW method. 
His calculations are in overall agreement with experiment,
but still remarkable discrepancies are found for the Ni [110] and
[111] MCP spectra.\\
The electronic structure of fcc Ni has been 
subject of intensive studies as a prototype of itinerant
 electron ferromagnets, since they indicate a failure of 
the one-electron theory \cite{L79, L81, WC77, EP80}. 
The LSDA calculations for fcc Ni cannot reproduce some 
features of the electronic structure of Ni observed experimentally. 
The valence band photoemission spectra of Ni \cite{EHK78, HKE79} 
shows a 3d-band width that is about 
30 \% narrower than obtained from the LSDA calculations \cite{WC77}. 
Second, the spectra show a dispersionless feature at about
 6 eV binding energy (the so-called 6 eV satellite) 
\cite{HW75, GBP+77}, which again cannot be reproduced by the LSDA calculations. 
 Third, the magnetic exchange-splitting is
 overestimated by LSDA calculations \cite{WC77} compared with the experimental
 data \cite{DGM78}. On the other hand, an improved description of correlation
 effects for the 3d electrons via the LSDA+DMFT \cite{LKK01, MCP+05, BME+06,
 Min11} 
gives the width of the occupied 3d bands of Ni properly and reproduce the exchange splitting and the 6 eV satellite structure in the valence band.\\
In view of these LSDA+DMFT improvements \cite{LKK01, MCP+05} 
upon the magnetic properties of Fe and Ni, the comparison of experimental 
MCP with the LSDA+DMFT theory provides some new information besides offering a
test for the impact of electronic correlations. In particular 
we demonstrate here that the LSDA+DMFT calculations improves 
also the agreement between theory and experiment for Fe and Ni MCPs. 
In the following we briefly discuss the theoretical approach and present the
 calculated MCPs of Fe and Ni together with the experimental data.
 Finally, the LSDA and LSDA+DMFT calculated MCPs of Ni have been 
decomposed and the contribution of different energy windows 
in the valence band have been compared in order to extract 
the features of electron correlations and to show their energy-dependency. 
Such a decomposition provides the evidence of the connection between 
the MCP contribution in the lower part of the valence band and the existence
of 6 eV satellite, both features being captured only within LSDA+DMFT. \\
The calculations were done using the spin-polarized relativistic
Korringa-Kohn-Rostoker (SPR-KKR) method in the atomic sphere approximation
(ASA) \cite{EKM11}. The computational scheme is based on the KKR 
Green function
formalism, which makes use of multiple scattering
theory, and was recently extended to compute MCPs \cite{SGST84,BME06,DB04}.
The spin projected momentum density $n_{m_s}(\vec p)$ (where $m_s=\uparrow
(\downarrow)$ ) is computed
using the LSDA(+DMFT) Green's functions in momentum space as
\begin{equation}\label{e7}
n_{m_s}(\vec p)={-\frac{1}{\pi} \int_{-\infty}^{E_F}
\Im G_{m_s}^{LSDA(+DMFT)}(\vec p,\vec p,E)dE}\,\nonumber.
\end{equation}
In order to analyze the momentum density and the corresponding MCPs in different energy ranges, we use a decomposition
of the above formula  in the form:
\begin{equation}\label{e8}
n_{m_s,\Delta E}(\vec p)={-\frac{1}{\pi} \int_{E_1}^{E_2}
\Im G_{m_s}^{LSDA(+DMFT)}(\vec p,\vec p,E)dE}\,\nonumber,
\end{equation}
where $ \Delta E = E_2 - E_1$  represent the width of the energy window.
The MCP seen in each energy window $\Delta E$
is obtained by performing a double integral in the momentum plane
perpendicular to the scattering momentum $\vec p_z$
\begin{equation}\label{mcs3}
J_{mag}^{LSDA(+DMFT)}(p_z)=
\int \int \left( n_{\uparrow}(\vec p) -
 n_{\downarrow}(\vec p) \right) dp_x dp_y\nonumber.
\end{equation}
Here the electron momentum density for a given spin orientation is
given by $n_{\uparrow (\downarrow)}(\vec p)$.
The area under the MCP is equal to the spin moment per
Wigner-Seitz cell:
$\int_{-\infty}^{+\infty} J_{mag}^{LSDA(+DMFT)}(p_z) dp_z =
\mu_{spin}^{LSDA
(+DMFT)}$.
In the actual calculations the experimental
lattice parameters of Fe and Ni have been used
($a_{Fe/Ni}=0.287/0.352$ nm). The exchange-correlation potentials parameterized by Vosko, Wilk and Nusair 
\cite{VWN80} were used for the LSDA calculations. For 
integration over the Brillouin zone the special points 
method has been used \cite{MP76}. 
In addition to the LSDA calculations, a charge and self-energy self-consistent
scheme for correlated systems based on the KKR approach with the 
many body effects 
described by the means of dynamical mean field theory (DMFT) has 
been applied \cite{MCP+05}.
As a DMFT solver the relativistic version of the so-called Spin-Polarized
T-Matrix Fluctuation Exchange approximation \cite{KL02, PKL05} was used. 
The realistic multi-orbital interaction has been parameterized by the average screened Coulomb interaction $U$ and the Hund
exchange interaction $J$. The $J$ parameter can be calculated directly 
within the LSDA and 
is approximately the same for all 3d elements. 
The parameter $U$ is strongly affected by the 
metallic screening 
and it is estimated for the 3d metals
to lie between 1 and 3 eV. In the present calculations we used
 $J$ = 0.9 eV and $U$ = 2.3 eV.\\
The MCPs of Ni [111] calculated on the basis 
of the LSDA and LSDA+DMFT, respectively, are shown in Fig. 
\ref{Fig:figure1}, together with the experimental data.
The Gaussian broadening applied to the calculated 
MCPs corresponds to the experimental resolution. 
The experimental MCPs stemming from Dixon 
et al. \cite{DDG+98} have been normalized to the experimentally 
determined spin moment (0.56 $\mu_B$). After broadening, the calculated KKR MCP spectra have been normalized to the calculated spin moment 
(0.6 $\mu_B$ by LSDA and LSDA+DMFT). \\
As can be seen, in the high-momentum region ($p_z \ge 2$ a.u.) the
correlation effects have a small influence on the magnetic spin density. 
In the momentum region $0 \le p_z \le 2$ a.u., taking into account the electron correlations by LSDA+DMFT approach improves the agreement with the experimental spectra considerably. Our LSDA+DMFT calculations can reproduce the dip in the [111] profile at $\sim$ 0.8 a.u., which was clearly underestimated by the LSDA calculations. \\
\begin{figure}[h]
\vspace{0.5cm}
\begin{center}
  \includegraphics[width=0.79\linewidth, clip=true]{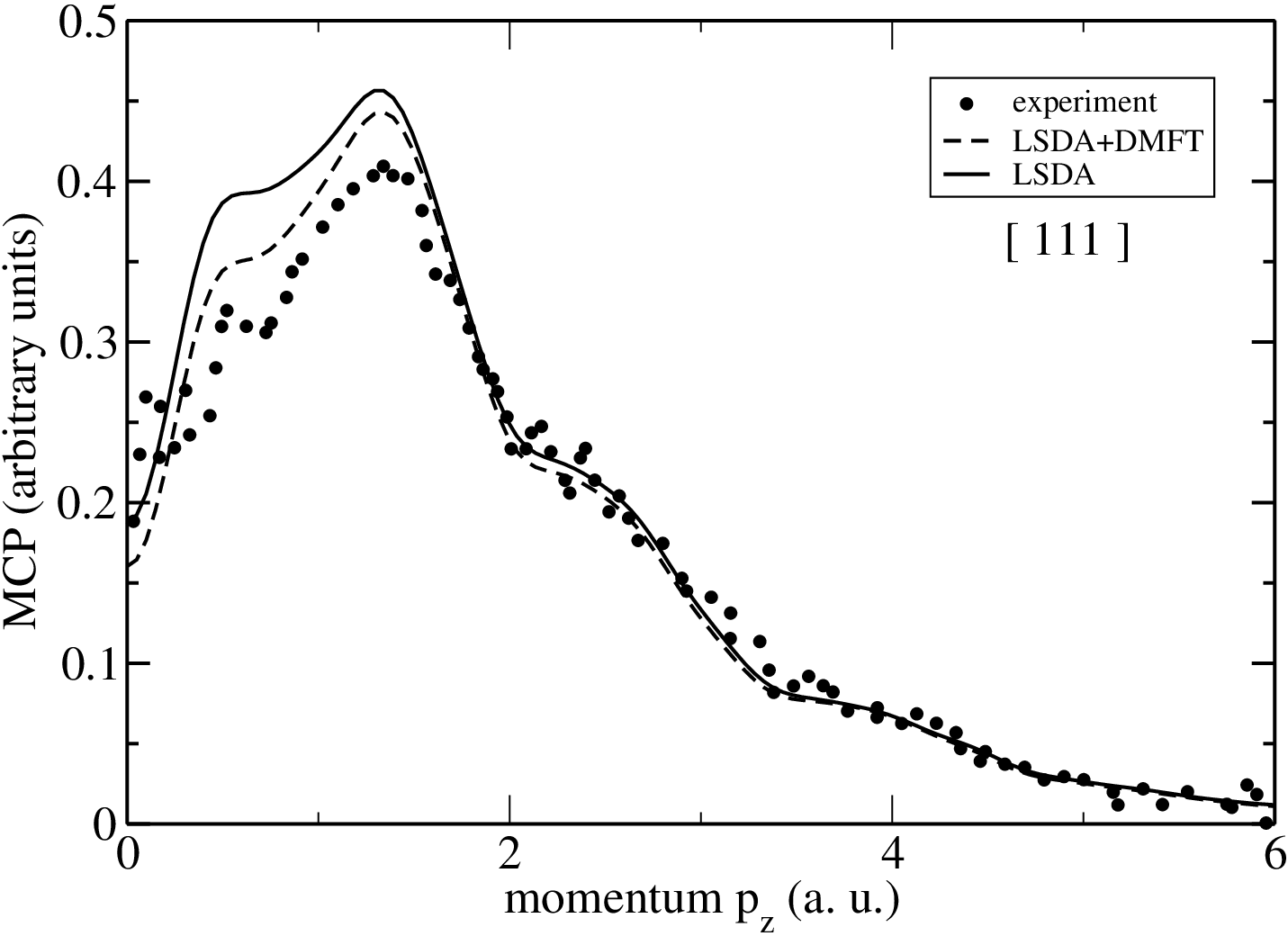}
\end{center}
\caption {\label{Fig:figure0} MCPs of Fe [111] calculated via the KKR method within the LSDA and LSDA+DMFT approach. 
The theoretical curves have been
convoluted with a Gaussian of 0.42 a.u. FWHM, corresponding to the experimental resolution. The experimental data stem from McCarthy 
et al.  \cite{McCCL+97}.}
\end{figure}
The LSDA and LSDA+DMFT calculated MCPs of Fe [111] are shown in Fig. \ref{Fig:figure0} together with the experimental spectra of McCarthy et al.  \cite{McCCL+97}. The experimental MCP has been normalized to a spin momentum of 2.07 $\mu_B$. The calculated spectra have been convoluted with a Gaussian of 0.42 a.u., corresponding to the experimental resolution. After convolution, the calculated MCPs have been scaled at a spin momentum of 2.3 $\mu_B$ (the LSDA calculated MCP) and 2.19$\mu_B$ (the LSDA+DMFT calculated MCP), respectively. Although both calculated MCPs show agreement with the experiment in the high momentum region, the shoulder at $\sim$ 0.5 a.u. is diminished in the LSDA+DMFT calculated MCP, improving the agreement with the experimental spectra also in the low momentum region. \\      
\begin{figure}[h]
\begin{center}
  \vspace*{0.4cm}
\hspace*{0.2cm}
 \includegraphics[width=0.79\linewidth,clip=true]{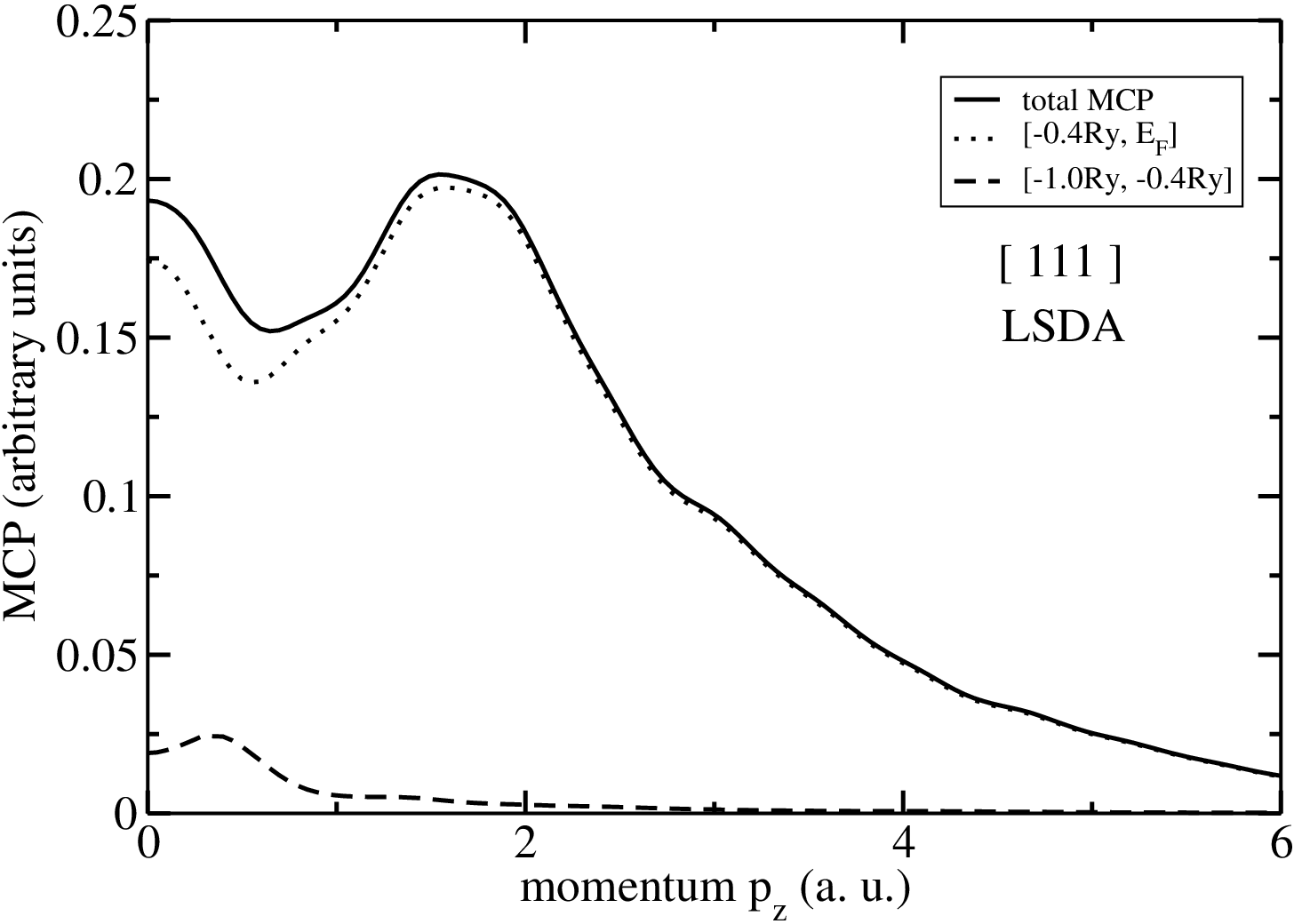}
\end{center}
\caption{\label{Fig:LSDA}The theoretical MCPs of Ni [111] obtained from LSDA calculations. The MCPs have been 
decomposed into 
contributions of two energy windows: [-1.0 Ry, -0.4 Ry] and [-0.4 Ry, E$_F$].}
\end{figure}
\begin{figure}[h]
\begin{center}
\hspace*{0.2cm}
 \includegraphics[width=0.79\linewidth,clip=true]{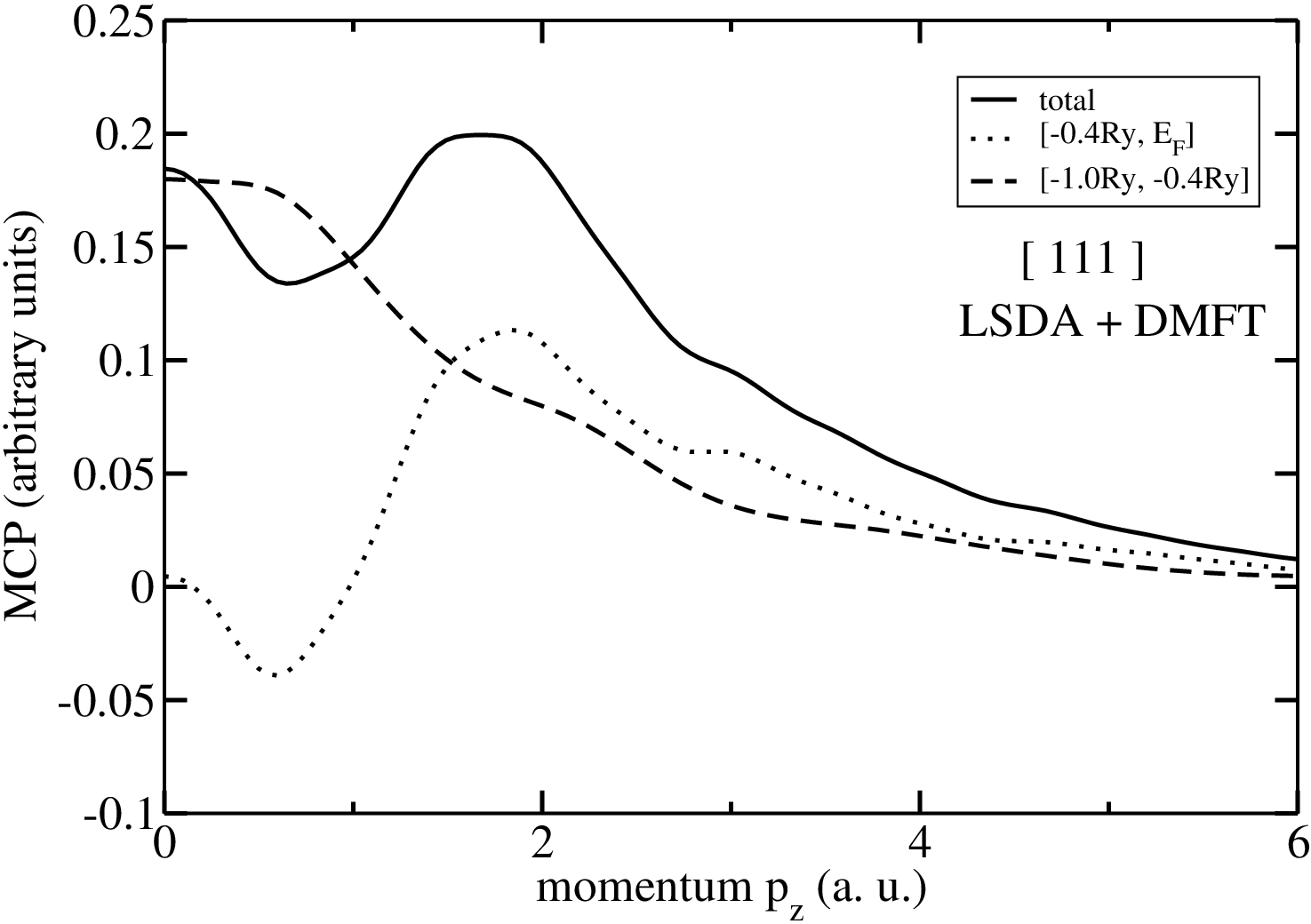}
\end{center}
\caption{\label{Fig:LSDA+DMFT}The theoretical MCP of Ni [111] obtained from 
KKR LSDA+DMFT calculations. The MCPs have been decomposed into contributions of two energy windows: [-1.0 Ry, -0.4 Ry] and [-0.4 Ry, E$_F$].}
\end{figure}
The theoretical MCP of Ni [111] has been decomposed into contributions
 from different energy windows. To illustrate the decomposition into
 energy windows, the LSDA calculated MCP stemming from the
 energy bands below the Fermi level in the range [-0.4 Ry, E$_F$] and
 [-1.0 Ry, -0.4 Ry]  are shown in Fig. \ref{Fig:LSDA}.
 As can be seen, the main contribution of the LSDA calculated MCP is stemming from the energy bands situated  between -0.4 Ry and the Fermi level. The bands at energy lower than -0.4 Ry have just a small positive contribution to
 the MCP within the momentum range $p_z \le 1$ a.u.. 
The corresponding decomposition has been performed as well for the LSDA+DMFT calculated MCP and the results are shown in Fig. \ref{Fig:LSDA+DMFT}. 
The contribution of the energy bands situated between -0.4 Ry and Fermi level has an important negative polarization for $p_z \le 1$ a.u.. According to the s, p and d-electron decomposition (not shown here) the negative contribution stem from s and p electrons. The overall negative polarization of s and p electrons is increased in the LSDA+DMFT MCP compared with the LSDA approach.\\
An essential feature is the important positive contribution of the electronic states in the energy window [-1.0 Ry, -0.4 Ry] to the MCP, that is similar for different scattering directions. 
The MCP carries information about all spin-polarized electrons in the system
 and about their localization. A broad contribution in the MCP is an
 indication for dominating localized spin states \cite{C85}. As the LSDA+DMFT MCPs  are broader than the LSDA ones we have a clear
indication that the spin magnetic densities have the tendency to localize in
the presence of the electronic correlations. As this MCP broadening is
seen in the energy range [-1.0 Ry, -0.4 Ry] where the well known feature of
the correlated electronic bands of Ni, namely the 6 eV satellite (0.44
Ry) is situated, a direct connection between this two correlation
features is presumable. As was shown by earlier
calculations 
\cite{LKK01} and was confirmed by photoemission
 experiments \cite{APM+00}, the 6 eV satellite is spin polarized and
 accordingly has to be connected with the MCP. The general 
interpretation of the 6 eV satellite relates this feature to an excited state
 involving two 3d holes bound on the same Ni site, therefore it is not 
accessible to any LSDA calculations. On contrary the LSDA+DMFT
 approach is able to capture such processes via the explicit existence within the 
interacting Hamiltonian of the four index form of the Coulomb matrix. Although the
 correlation induced satellite is absent in Fe, the proper description of the 
 angle-resolved photoemission spectra cannot be done by LSDA but by LSDA+DMFT
 \cite{SBF+09}. In the case of Fe, a broad contribution of the LSDA+DMFT MCP
 is also obtained in the energy range [-1.0 Ry, -0.4 Ry] and consequently 
a similar tendency of localization of the spin density is expected.\\
In conclusion, the MCPs of Fe and Ni have been determined using the SPR-KKR band structure 
method within  the LSDA and LSDA+DMFT approach, respectively. 
The influence of electron correlations on the MCP of Fe and Ni 
[111] has been discussed. For high transfer momenta (($p_z \ge 2$ a.u.) 
no significant corrections due to dynamical electronic correlations 
is seen in the Compton profile, therefore 
the dynamics of the process can be captured equally well by a LSDA approach. 
On contrary for small momentum transfer, a clear evidence for the 
interplay between the energy transfered to electron and the
 electronic correlations is seen: including the local but 
dynamical self-energy leads to  an improved MCP spectra. 
In addition the decomposition of the Ni [111] 
MCPs shows a large and broad contribution by DMFT+LSDA 
approach stemming from the energy window between the bottom 
of the valence band and 0.4 Ry binding energy. For the
 corresponding MCP decomposition the LSDA approach shows 
just a small and narrow contribution stemming from the same energy range.  
We consider this feature as a consequence of the localization tendency
of the spin density due to electronic correlations.\\


DB acknowledge the financial support of the
Deutsches Forschungsgemeinschaft through FOR 1346 and the 
project POSDRU 89/1.5/S/60189  with the title 
'Postdoctoral Programs for sustainable Development in a Knowledge 
Based Society'. Financial support by
 the Bundesministerium f\"{u}r Bildung und Forschung 
(05K10WMA) is also gratefully acknowledged.
    

\bibliographystyle{apsrev}


\end{document}